\documentclass[conference]{IEEEtran}
\usepackage{blindtext, graphicx}
\usepackage[dvipsnames]{xcolor}
\usepackage{subcaption}
\usepackage{wrapfig}
\usepackage{amsmath}
\usepackage{textcmds}
\usepackage{amssymb}
\usepackage{algorithm2e}
\usepackage{booktabs}
\usepackage[letterpaper, left=0.64in, right=0.64in, bottom=1.05in, top=0.75in]{geometry}

\SetKwComment{Comment}{/* }{ */}
\makeatletter
\def\ps@IEEEtitlepagestyle{%
  \def\@oddfoot{\mycopyrightnotice}%
  \def\@oddhead{\hbox{}\@IEEEheaderstyle\leftmark\hfil\thepage}\relax
  \def\@evenhead{\@IEEEheaderstyle\thepage\hfil\leftmark\hbox{}}\relax
  \def\@evenfoot{}%
}
\def\mycopyrightnotice{%
  \begin{minipage}{\textwidth}
  \centering \scriptsize
  Copyright~\copyright~20xx IEEE. Personal use of this material is permitted. Permission from IEEE must be obtained for all other uses, in any current or future media, including\\reprinting/republishing this material for advertising or promotional purposes, creating new collective works, for resale or redistribution to servers or lists, or reuse of any copyrighted component of this work in other works by sending a request to pubs-permissions@ieee.org.
  \end{minipage}
}
\makeatother
\begin{document}
% Use-Cases for Connected and Autonomous Vehicles: V2X Communication Architecture, Requirements and Design Considerations

\title{Multi-Task Learning as enabler for General-Purpose AI-native RAN}

\author{\IEEEauthorblockN{Hasan Farooq, Julien Forgeat, Shruti Bothe, Kristijonas Cyras and Md Moin}
\IEEEauthorblockA{Ericsson Research, Santa Clara, 95054 CA, USA \\
\{hasan.farooq, julien.forgeat, shruti.bothe, kristijonas.cyras, md.moin.uddin.chowdhury\}@ericsson.com}
}
\maketitle

\begin{abstract}
The realization of data-driven AI-native architecture envisioned for 6G and beyond networks can eventually lead to multiple machine learning (ML) workloads distributed at the network edges driving downstream tasks like secondary carrier prediction, positioning, channel prediction etc. The independent life-cycle management of these edge-distributed independent multiple workloads sharing a resource-constrained compute node e.g., base station (BS) is a challenge that will scale with denser deployments. This study explores the effectiveness of multi-task learning (MTL) approaches in facilitating a general-purpose AI-native Radio Access Network (RAN). The investigation focuses on four RAN tasks:  (i) secondary carrier prediction, (ii) user location prediction, (iii) indoor link classification, and (iv) line -of-sight link classification. We validate the performance using realistic simulations considering multi-faceted design aspects of MTL including model architecture, loss and gradient balancing strategies, distributed learning topology, data sparsity and task groupings. The quantification and insights from simulations reveal that for the four RAN tasks considered (i) adoption of customized gate control-based expert architecture with uncertainty-based weighting makes MTL perform either best among all or at par with single task learning (STL) (ii) LoS classification task in MTL setting helps other tasks but its own performance is degraded (iii) for sparse training data, training a single global MTL model is helpful but MTL performance is on par with STL (iv) optimal set of group pairing exists for each task and (v) partial federation is much better than full model federation in MTL setting.
\end{abstract}


\IEEEpeerreviewmaketitle

\section{Introduction}

AI-native RAN is a trending concept that is spearheading the evolution of wireless mobile networks. This concept makes AI pervasive in the entire RAN architecture. The reason being AI’s pronounced success for virtually every RAN design/optimization aspect \cite{7982603}. These AI driven RAN workloads can be deployed at the edge of networks e.g., centralized unit (CU)/distributed unit (DU). Edge-deployed ML workloads, in contrast to traditional cloud-centric architecture, are appealing for fulfilling the latency, scalability, reliability, and privacy needs of beyond-5G applications. Additionally, they present an attractive solution for privacy-focused multi-vendor deployment scenarios and data regulatory compliance. 

Nevertheless, a notable limitation in the majority of AI-driven RAN studies is the tendency to address specific RAN problems, features, or use cases in isolation. In such cases, AI is often customized to suit a particular use case, referred to as specialized AI models. Implementing solutions with specialized AI models in live RANs may result in an uncontrolled proliferation of specialized ML models per downstream task (radio feature) that will increase both RAN complexity and operational expenditure.
In particular, the independent life-cycle management (LCM) of these edge-distributed independent multiple workloads sharing a resource-constrained compute node (BS, CU/DU) will be challenging in terms of availability of labelled data and compute-memory resources that will scale with denser deployments. Contrary to this, general purpose AI-native RAN vision is much more desirable wherein a single AI algorithm would have the capability to learn and manage a wide spectrum of networking operations, spanning the whole protocol stack. Achieving this involves designing an AI algorithm that can concurrently control multiple RAN tasks (Fig. 1). 
\begin{figure}[tb]
	\centering
	\includegraphics[height=0.12\paperheight]{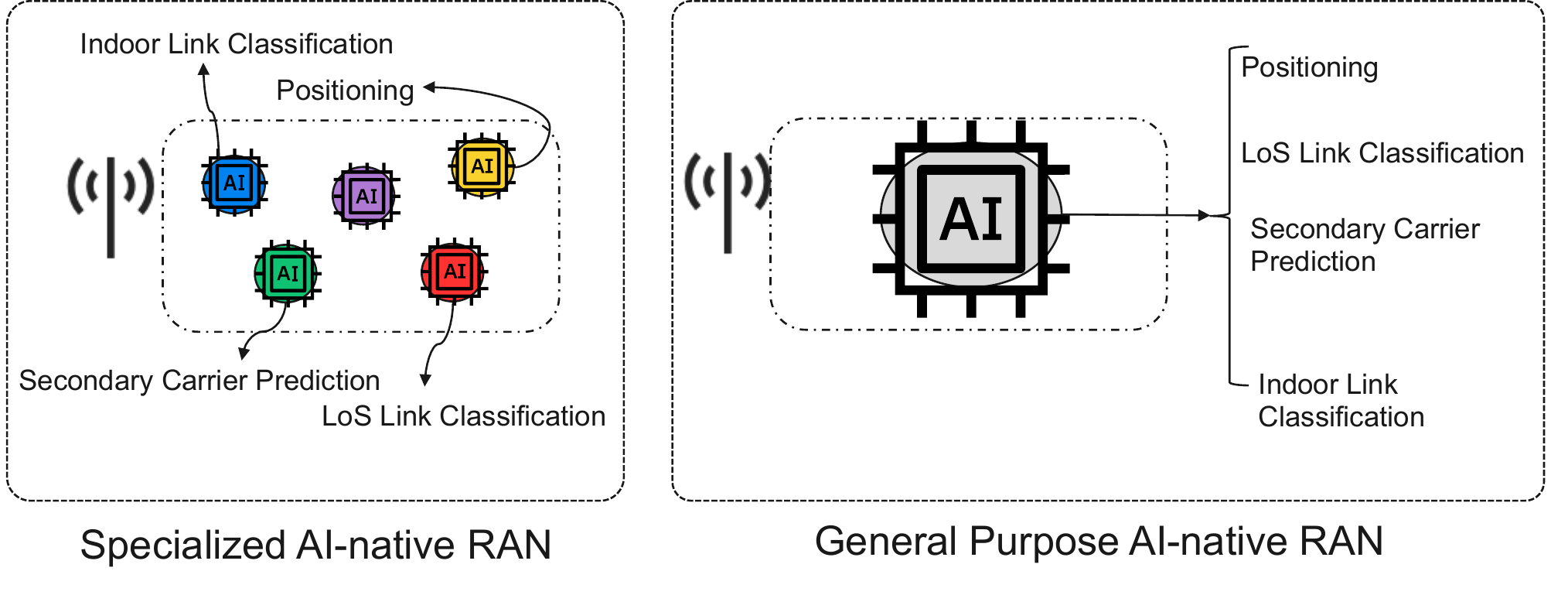}
	\caption{Specialized AI-native RAN vs General AI-native RAN.}
	\vspace{-7 mm}
\end{figure}

Multi-task learning (MTL) is one such paradigm that can be used to train a ML model to perform multiple RAN related tasks. MTL jointly learns multiple related tasks using a single model. MTL draws inspiration from human learning, where individuals frequently leverage knowledge acquired from prior tasks to facilitate the learning of a new task. MTL has been shown to improve parameter-efficiency and inference speed compared with learning a separate model for each task. It can also enable features for one task to be available for another and emphasizes generalizable features that are common across tasks. With an MTL approach, different tasks need to share some common structure otherwise it is usually better to use single-task learning. Fortunately, in the context of RAN, there are many tasks with shared structure that may also need concurrent execution. Even if tasks are seemingly unrelated, the laws of physics that govern radio propagation are same for all tasks. In this study, we aim to address three key research questions: \textit{ (i) What is the impact of the design space of state-of-the-art MTL approaches in terms of model architecture and weighting strategies on the overall performance of RAN-related tasks?  (ii) In various scenarios, which among the three modes of distributed learning topology—local model, global model, or partial federated model—is most appropriate? (iii) How does the sparsity of training data and the different groupings of RAN tasks influence the performance of MTL?}

\subsection{Related Work}
Some works have used ML for RAN tasks considered in this paper, such as \cite{8369000} for secondary carrier prediction, \cite{9417278} for location prediction, \cite{9954290} for indoor/outdoor link classification, and \cite{9569316} for LoS link classification. However, the conventional approach in these AI for RAN studies is to design a specialized independent ML (STL) model for each task. As opposed to STL, MTL is a promising paradigm that has been largely studied in the domain of computer vision and natural language processing related use-cases only as evident from survey in \cite{9392366}. Recently very few works have explored MTL paradigm in context of mobile networks like in \cite{9627159} which has explored MTL to address two RAN use cases, handover management and initial modulation selection. Likewise the work in \cite{9144137} used MTL for predicting user environment (indoor or outdoor) and mobility (low, medium or high) state using drive test data. The work in \cite{9500447} has used MTL to simultaneously learn modulation and signal classification tasks. All of these studies have used neural network (NN) based hard parameter sharing architecture without accounting for imbalanced losses or distributed learning topology that can have substantial effect on learning performance. To the best of authors’ knowledge, performance analysis of state-of-the-art MTL approaches for RAN tasks considering its key design aspects like model architecture, loss balancing, distributed learning topology is missing in literature. This paper aims to fill this gap.
\subsection{Contributions}
The contributions of this paper can be summarized as follows:

\begin{enumerate}
	\item 
	 We demonstrate how to design MTL approach to improve RAN related task performance by performing simulations to evaluate the behavior of four RAN tasks in multi-task learning context considering MTL model architecture, loss and gradient balancing strategies, and data sparsity.
	 \item
	We demonstrate how MTL performance varies when training independent MTL models, a single global MTL model and partial federated MTL model with two variants depending upon the order in which local parameters are updated.
	\item
	We present experimental results on how different grouping combinations of tasks affect overall MTL performance.
\end{enumerate}

The rest of the paper is organized as follows: Section II describes the MTL paradigm in context of four RAN use-cases. The performance analysis and associated discussion is presented in Section III while Section IV concludes the paper. 

\section{Multi-Task learning for RAN distributed workloads}

In the context of $T$ learning tasks, each having their own dataset $D_t = {(x,y)_t}$ where $x$ are input features, $y$ are labels and $L$ is the model loss. Each task is characterized by a tuple $\{p(x)_t, p(y|x)_t, L_t\}$ and each task should differ in at least one of them. For the scope of this paper, we have considered four ML driven RAN tasks running in each of the base stations:
\begin{enumerate}
	\item 
	 Secondary Carrier Prediction (SC): user equipments (UEs) use AI model to predict coverage on the higher band based on measurements at the serving lower band carrier. This avoids the need to perform measurements on a secondary carrier, thus reducing the energy consumption, throughput degradation and the delay.
	 \item
	Positioning (PS): location estimation/prediction enables advanced Location-based Services (LBS) and is also helpful for emergency (911) scenarios. It is also helpful in beam management (e.g., reducing search space of initial beam search procedure) and thus, improves the quality of service experienced by users.
	\item
	Indoor/Outdoor link Classification (IN): It makes it possible to exploit the environmental context to enhance network operations, in terms of better QoE e.g., handover to indoor small cells, slice selection etc.
	\item
	LoS/NLoS link Classification (LOS): identification of LoS/NLoS links in higher bands e.g., mmWave communication is important for range-based localization since utilizing NLoS measurements directly as LoS measurements may result in large localization errors.

\end{enumerate}

We have framed the MTL problem as follows: given the reference signal received power (RSRP) on LTE carrier for all cells, the objective is to predict: (a) RSRP on NR carrier for all cells (b) distance to BSs (c), classify if UE is indoor and (d) classify if user equipment (UE) has LoS link on NR carrier for all cells. In real networks, this dataset can be built by BSs using 3GPP standardized Minimization of Drive Test reporting (MDT) feature. BSs receive MDT reports from UEs that have information about their measured RSRPs for both intra and inter-frequency bands carriers (input features, secondary carrier prediction labels) and using it together with geographical datasets containing 3D Earth terrain and building layouts of the city to capture labels for location prediction and LoS link classifications tasks. Once the MTL model is trained, it can run in BSs requiring only intra-frequency radio measurements from UEs and inferring labels for the four tasks to be consumed in downstream tasks e.g., instructing UE to handover to a specific beam. In our scenario, the input features distribution would be the same across all tasks, as the RSRP levels for intra-frequency cells serve as the input features for each task. However, there is variation in the conditional distribution of labels given input features across tasks, as the labels differ among them. Loss function is also different across the tasks. 

Let each task $t$ has a training dataset $D_t$, MTL aims to learn a model on $D_t$ to learn $T$ tasks together to improve the learning of a model for each task by using knowledge in all or some of other tasks.  The most common way of doing MTL is through hard parameter sharing wherein NN parameters are split into task-sharing parameter $\theta^{sh}$ and task-specific parameters $\theta^{tsk}$. Let $L_t (D_t;\theta^{sh},\theta^{tsk})$ denote the average loss on $D_t$ for task $t$ using $(\theta^{sh},\theta^{tsk} )$. The objective function of MTL is: 
\begin{equation}
	\min_{\theta^{sh}, \theta^{tsk}}  \sum_{t = 1}^{T} \omega_t L_t (D_t;\theta^{sh}, \theta^{tsk})
\end{equation}
where $\omega_t$ is the task weight for task $t$. It makes sense to share some network structures, so each task doesn’t need to learn everything from scratch independently. While hard parameter sharing (HPS) \cite{HPS} serves as the foundational and frequently utilized structure in MTL, it can encounter challenges with negative transfer. This is attributed to potential conflicts among tasks, as parameters are shared directly. Some tasks have loss functions on different scales (classification vs regression), Some tasks are asymmetric in terms of importance, difficulty, data availability or noise.

To cope with these challenges, we formulated the research problem as follows: How to design MTL approach to improve RAN task performances by analyzing the state-of-the-art MTL approaches characterized by a tuple of \{architecture, weighting, distributed learning topology\}. Searching for an optimal combination of all these elements of tuple (optimization variables) with all of their possible search space is prohibitively expensive. Therefore for holistic optimization of MTL, we resorted to heuristic strategy by first exhaustively searching all possible combinations of architecture and weighting keeping distributed learning topology fixed. With the best performing combination of architecture and weighting, all possible distributed learning topologies are then evaluated in context of four RAN tasks. The search space for these elements used in experiments are described in next subsections.

\subsection{Architecture}
To deal with task conflicts we investigate three additional MTL model architectures (a) MMoE \cite{MMoE}, (b) Deselect-k \cite{Dselectk} and (c) CGC \cite{CGC} (see Fig. 2). In MMoE, the approach differs from having a single shared bottom network for all tasks. Instead, it employs a set of bottom networks, referred to as experts, along with a gating network for each task. The gating networks learns to assign weights to the experts on a per-example basis, and MMoE produces an output as a weighted combination of these experts. This per-example weighting mechanism enables distinct tasks to leverage the experts in varying ways making it more flexible and capable of capturing intricate relationships among tasks.

Dselect-k as opposed to MMoE uses sparse gates. CGC differs from MMoE in that it incorporates both shared and task-specific experts. Each expert module within CGC consists of multiple sub-networks referred to as experts. Shared experts in CGC focus on learning patterns that are common across tasks, while task-specific experts are responsible for extracting patterns specific to individual tasks.
\begin{figure}[tb]
	\centering
	\includegraphics[height=0.14\paperheight]{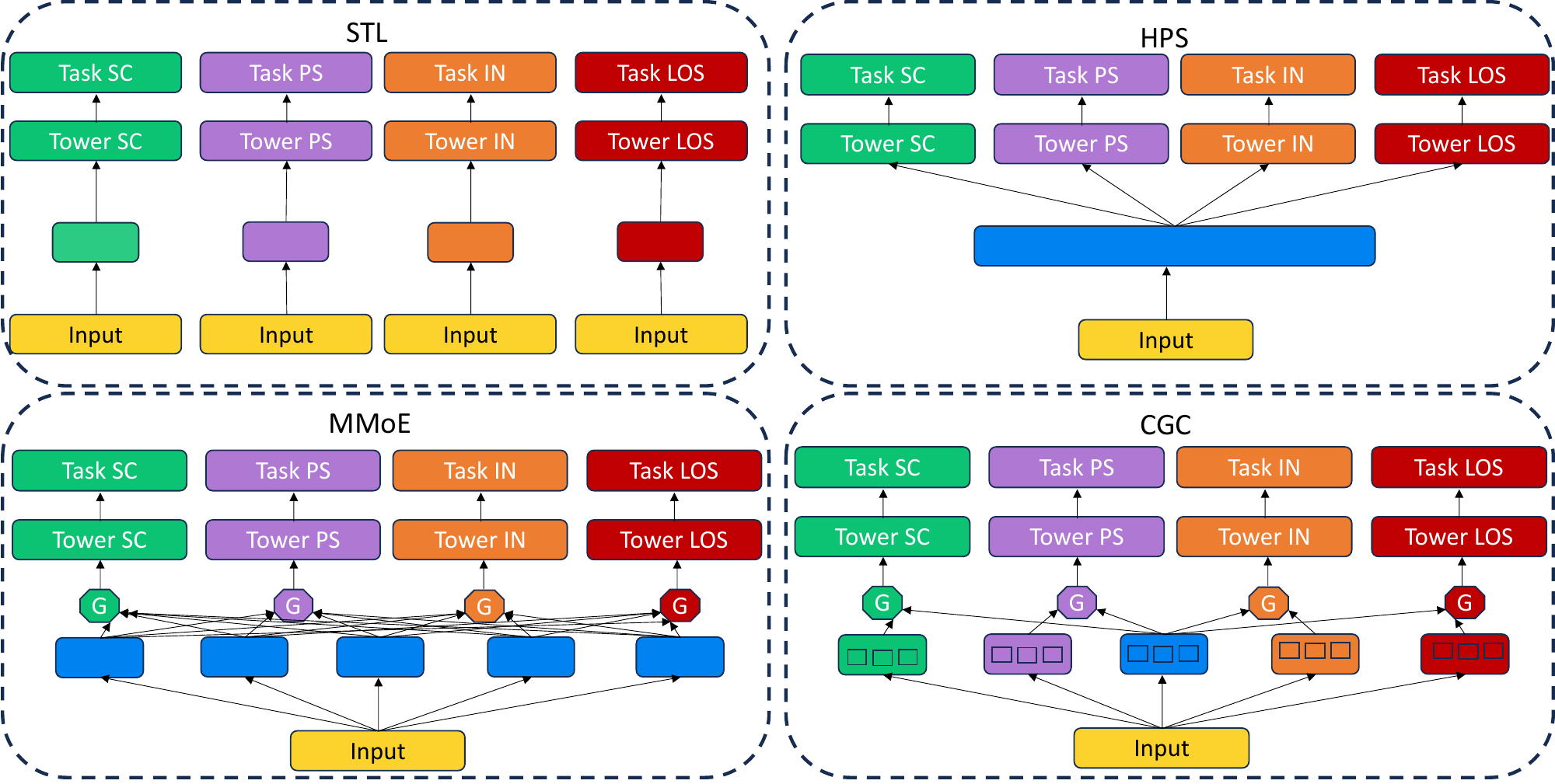}
	\caption{MTL Architectures. Here blue box shows shared parameters and nodes named 'G' are gates.}
	\vspace{-2 mm}
\end{figure}

\subsection{Weighting}
In addition to architecture which decides which parameters are shared and how to share, the other research thrust is devising strategies to optimize MTL. Balancing multiple training losses in MTL affects how task-shared parameters are updated and so methods like (i) loss balancing and (ii) gradient balancing are developed. Although, these two lines of research have been pursued independently in literature as the optimization methods are mainly related to the objective function, while the design of the architecture is to learn relationships between tasks. We in this work investigate loss balancing and gradient balancing schemes combined with architectures to cope with task negative correlation and further enhance MTL's performance. Instead of equally weighting (EW) all tasks simultaneously, loss balancing methods aim at weighting task losses computed dynamically according to different measures such as homoscedastic uncertainty (UW) \cite{8578879}. DWA \cite{8954221} learns to dynamically average task weighting over time by assessing the rate of loss change for each task. GLS in \cite{9025330} expresses the total loss of a multi-task learning problem as the geometric mean of individual task losses while RLW \cite{lin2022reasonable} uses normalized random weights for the tasks. 

From the gradient perspective, the update of task-sharing parameters $\theta$ depends on all task gradients. Thus, gradient balancing methods aim to aggregate all task gradients under different constraints. For example, MGDA \cite{MGDA} formulates MTL as a multi-objective optimization problem and directly solves the optimal weights in every iteration by finding a common descending direction among all the gradients via solving a quadratic programming problem. CAGrad \cite{CAGrad} improves MGDA by constraining the aggregated gradient to around the average gradient. GradNorm \cite{GradNorm} learns task weights by constraining the gradient magnitude of each task to be similar. PCGrad \cite{pcgrad} projects the gradient of one task onto the normal plane of the others if their gradients conflict while GradVac \cite{wang2021gradient} aligns the gradients regardless of whether the gradients conflict or not. GradDrop (\cite{gradrop} randomly masks out the gradient values with inconsistent signs.

\subsection{Distributed Learning Topology}
We also consider distributed learning mode depending upon where training of MTL models happen within RAN context. (a) local mode (MTL\_local, STL\_local) wherein each model will reside in BS and have its own data and will train model independently without sharing anything with centralized server, (b) global mode (MTL\_global, STL\_global) wherein all sites’ data is aggregated into a global dataset at a centralized place and a single global model is trained for all BSs, and (c) partial federated approach wherein only the weights of shared layers are used in federation. In this setting, eq (1) modifies to (2):  
\vspace{-2 mm}
\begin{equation}
	\min_{\theta^{sh}, \{\theta^{tsk}_i\}_{i=1}^{N_{BS}}} \frac{1}{N_{BS}} \sum_{i = 1}^{N_{BS}} \sum_{t = 1}^{T} \omega_{i,t} L_{i,t} (D_{i,t};\theta^{sh}, \theta^{tsk}_i)
\end{equation}

\noindent where $N_{BS}$ is the total number of BSs. During each communication round (k = 1, 2, ..., K) of partial federation, the server broadcasts the current global version of the shared parameters (e.g., shared\_experts in CGC) to participants BSs. Each BS uses the global shared representation part $\theta^{sh}$ received from the server to join with the task-specific parameters $\theta^{tsk}_i$ maintained locally to obtain the local model. It then performs one or more steps of (stochastic) gradient descent to update both the shared parameters and the personal parameters (e.g., task specific experts in CGC) and sends only the updated shared parameters to the server for aggregation (Fig. 3). We explored two partial federation algorithms FedAlt and FedSim \cite{Pillutla2022FederatedLW} that differ in way how local updates are performed:
\begin{itemize}
  \item FedAlt: the BSs first update the personal parameters with the received shared parameters fixed for $\tau_{tsk}$ head local iterations and then update the shared parameters with the new personal parameters fixed for $\tau_{sh}$ local iterations.
  \item FedSim: the BS makes $\tau_{sh, tsk}$ local gradient-based updates to both its shared and the task-specific layers synchronously.
\end{itemize}
In the beginning, we let the BSs run their MTL locally for few epochs ($\tau_{inital}$)  and then start federation rounds that showed to improve performance as compared to starting federation from start like in FedVanila benchmark we used in our experiments. We explored hyper-parameters for FedAlt and FedSim and found that FedAlt is more sensitive to its hyper-parameter values. For FedSim best parameters were \{$\tau_{inital} = 10, \tau_{sh, tsk} = 10, K = 9$\} while for FedAlt, \{$\tau_{inital} = 10, \tau_{tsk} = 17, \tau_{sh} = 1, K = 5$\}. For FedAlt, keeping $\tau_{sh} $ smaller than $\tau_{tsk} $ gave better performance. We also experimented by sharing task specific parameters of CGC in federation rounds while keeping shared experts out but the performance degraded. We also compared performance with FedVanila in which all model parameters are averaged by server.
\begin{figure}[tb]
	\centering
	\includegraphics[height=0.15\paperheight]{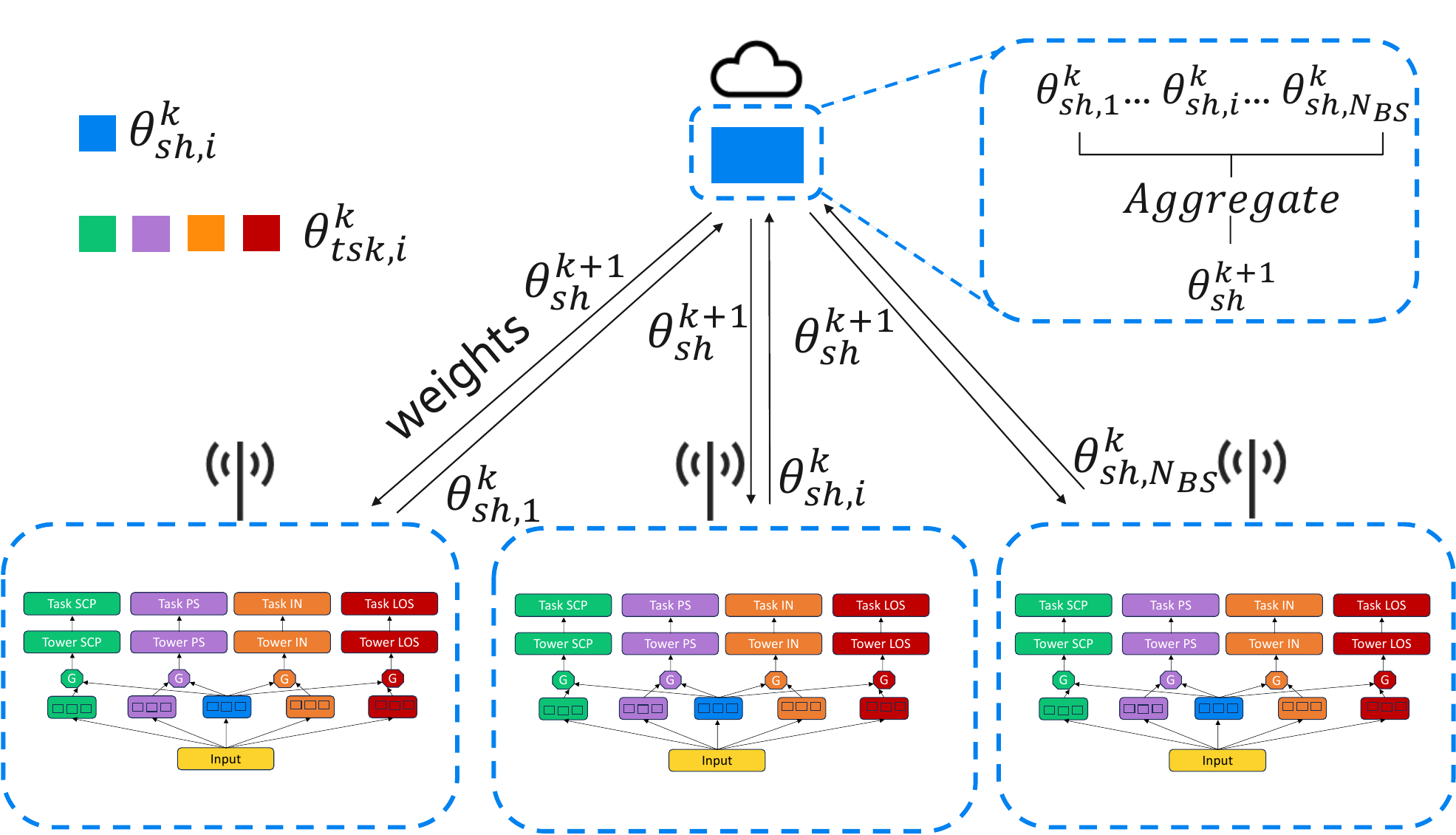}
	\caption{Partial Federated MTL. }
	\vspace{-2 mm}
\end{figure}

\section{Results and Discussion}\label{sec:model}

We generated the dataset from Ericsson propriety radio network simulator with propriety 3D ray tracing propagation model. We generated data from multiple city scenarios each having three base stations. Each base station has three sectors supporting one primary LTE and one secondary NR carrier. The simulated area consisted of a 2 km x 2 km slice of large cities from Europe, Asia and USA containing several buildings of varying heights causing scattering effects. Each base station data was divided into training/validation and testing with ratio 60/20/20. The input features are intra-RSRP measurements of all cells that will be same for all RAN tasks. The task labels are:
\begin{enumerate}
	\item 
	 Secondary Carrier Prediction (SC): Measured inter-RSRP on secondary frequencies for all cells with dimension $1 \times N^{fs}_{cells}$.
	 \item
	Positioning (PS): Distance of UE to all BSs with dimension $1 \times N_{BS}$.
	\item
	Indoor/Outdoor link Classification (IN): Binary scalar indicator if UE is indoor.
	\item
	LoS/NLoS link Classification (LOS): Binary indicator if UE has LoS link for each of the secondary carrier cells with dimension $1 \times N^{fs}_{cells}$.

\end{enumerate}

\noindent where $N^{fs}_{cells}$ is the number of secondary carrier cells. The input features and output task labels for each of the BSs of a city comprises of cell measurements for only that city. For the MTL architectures and weighting strategies, we used the implementations available in python open-source library LibMTL \cite{LibMTL}. The network simulation parameters are given in Table I. Mean Squared Error (MSE) loss was used for SC and PS with mean absolute error (MAE) as their validation metric while binary cross entropy loss was used for IN and LOS tasks with accuracy as their validation metric. For MTL model training, we set training budget in terms of 100 epochs and we used Adam optimizer (lr = 0.0001, momentum = 0.9), 512 layer size for shared parameters, batch size of 64 and number of experts to be 2. Each expert was single layer feed-forward network. Rest of the parameters used were set to default values as provided by \cite{LibMTL}.

First, we explored the MTL design search space with distributed learning topology fixed to local mode. Each experiment is repeated multiple times and averaged results among all BSs across all four cities are reported. The box plot depicted in Fig. 4 illustrates the variability in values for the objective function $\Omega = Accuracy_{IN} + Accuracy_{LOS} - MAE_{SC} - MAE_{PS}$ across all conceivable combinations of MTL model architectures and weighting strategies. In this context, higher values indicate better performance. Based on lower 25 percentile, it appears that CGC with UW and HPS-GLS perform best among all. CGC with UW had better convergence and therefore this combination was chosen for rest of experiments. 
Next, we evaluate the performance for all four RAN tasks considering all modes of distributed learning and performance is shown in Figs. (5-9) for held-out test set. The inset bar plot shows percentage improvement as compared to using STL in local mode.
\begin{table}[b]
	\centering
	\caption{Network Simulation Parameters.}
	\begin{tabular}[t]{cc}
		\hline
		\multicolumn{1}{|c|}{\textbf{Parameter}}  & \multicolumn{1}{c|}{\textbf{Value}} \\ \hline
		\multicolumn{1}{|c|}{Num of cities}          & \multicolumn{1}{c|}{4}           \\ \hline
		\multicolumn{1}{|c|}{Num of BSs}          & \multicolumn{1}{c|}{3 per city}         \\ \hline
		\multicolumn{1}{|c|}{Propagation model} & \multicolumn{1}{c|}{3D ray tracing}     \\ \hline
		\multicolumn{1}{|c|}{LTE carrier}  & \multicolumn{1}{c|}{900 MHz}             \\ \hline
		\multicolumn{1}{|c|}{NR carrier}  & \multicolumn{1}{c|}{4.5 GHz}             \\ \hline
		\multicolumn{1}{|c|}{UE density} & \multicolumn{1}{c|}{1K uniformly distributed UEs per snapshot}             \\ \hline
		\multicolumn{1}{|c|}{Num of snapshots} & \multicolumn{1}{c|}{350 for four cities}             \\ \hline
		\multicolumn{1}{l}{}                      & \multicolumn{1}{l}{}               
	
	\end{tabular}
	%\vspace{-5 mm}
\end{table}

\begin{figure*}[tb]
	\centering
	\includegraphics[height=0.22\paperheight]{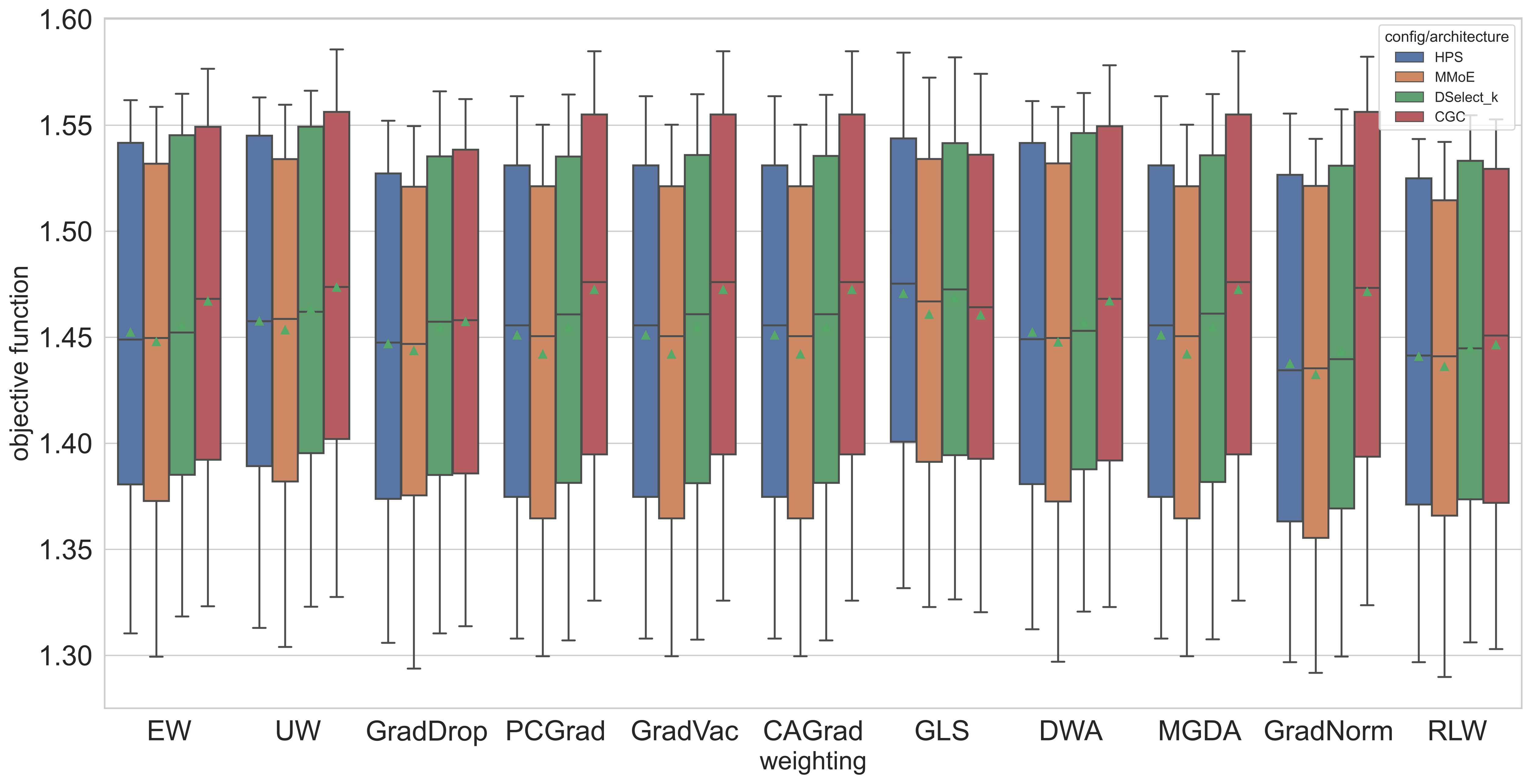}
	\caption{Effect of MTL architectures and weighting strategies.}
	\vspace{-5 mm}
\end{figure*}

\begin{figure}[tb]
	\centering
	\includegraphics[height=0.21\paperheight]{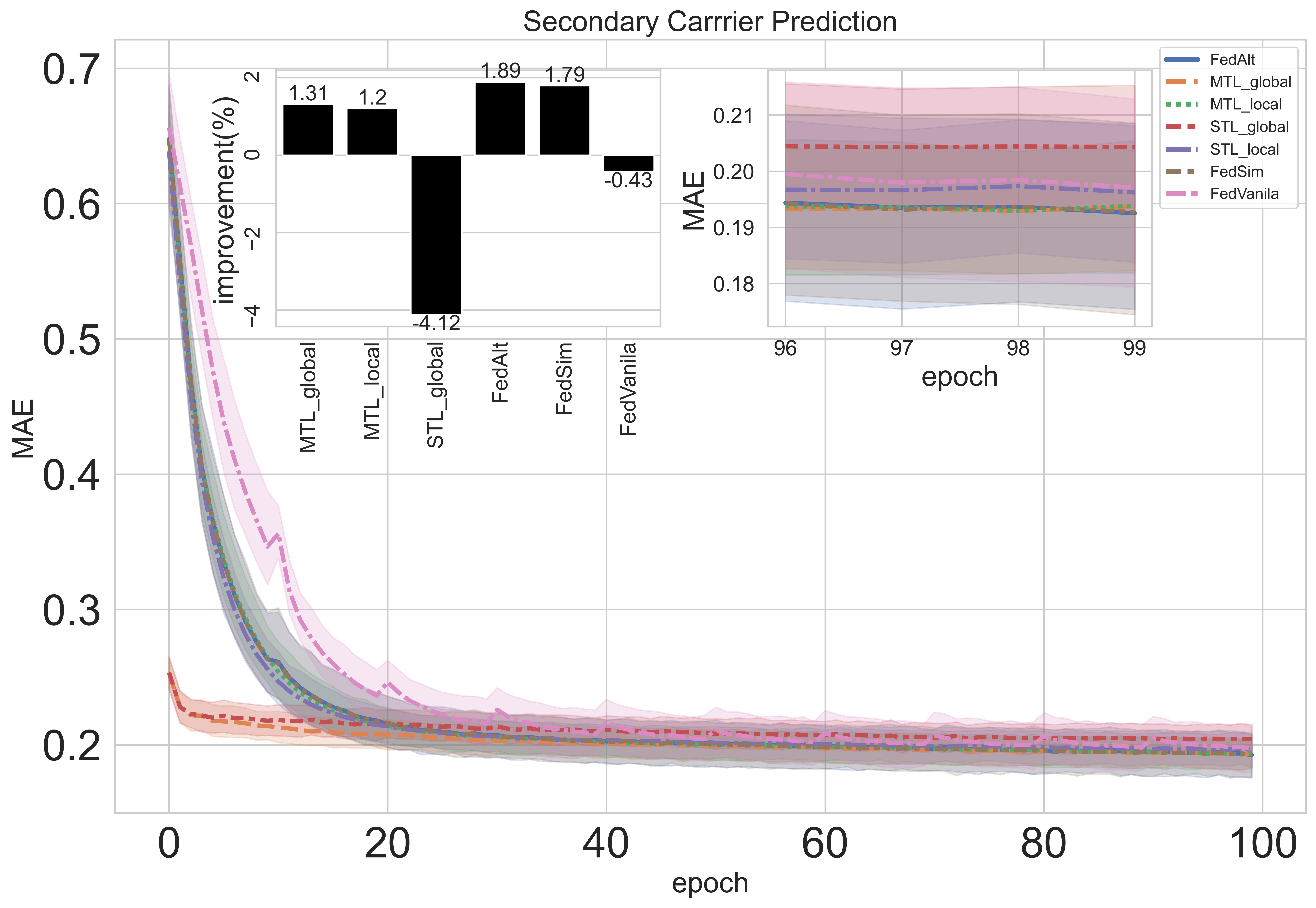}
	\caption{Secondary Carrier prediction task performance. Improvement over STL\_local = \{MTL\_global: 1.31\%, MTL\_local: 1.2\%, STL\_global: -4.12\%, FedAlt: 1.89\%, FedSim: 1.79\%, FedVanila: -0.43\%\}}
	\vspace{-5 mm}
\end{figure}

For secondary carrier prediction task with MAE as metric (Fig. 5), FedAlt (1.9\% improvement as compared to STL local) and FedSim (1.8\% improvement as compared to STL local) outperformed all others then comes MTL global and MTL\_local. The performance of STL\_global is worst. Both global based approaches (MTL\_global and STL\_global) have faster convergence while FedVanila has slowest convergence. In STL, the difference between global and local is much larger than in MTL local and global. Both global and local based MTLs performed better than STLs. For location prediction task with MAE as metric (Fig. 6), MTL\_global showed best convergence and outperformed all by large margins (9.6\% improvement) and later followed by FedSim and FedAlt. For both MTLs and STLs, global performed better than local. FedVanila has slowest convergence and STL\_local is worst.  Difference between STL\_global and local is much larger than MTL\_local and global. Overall globals are better than locals and MTL is better than STL. For indoor link classification with accuracy as metric (Fig. 7), MTL\_local is slightly better by STL\_local (0.05\% improvement). This is contrary to previous two where globals performed better than locals. Both MTL\_global and STL\_global have quick convergence to suboptimal value and at end STL\_global and MTL\_global are worst. For LoS link classification task and accuracy as metric (Fig. 8), STL\_local is best followed closely by MTL\_local. This is different from previous three where MTL performed best than STLs. FedVanila has unstable progression and slowest convergence. Both locals are better than global while MTL and STL are almost on par.

Overall, MTL performs either best among all or on par with STL. This is still an improvement as LCM costs are reduced i.e., one model instead of managing multiple models. Federation with partial model is better than with full model parameters. For secondary carrier prediction, collaboration through partial federation and for location prediction, collaboration through aggregating training data for global model or sharing model weights in partial federation is helpful. Global based approach works best for location prediction and for secondary carrier prediction, it is either on par with local (MTL\_local) or show degraded performance (STL\_global) while both classification tasks are better in local modes. One possible explanation for this can be the influence of local context. For location prediction, distance to base station is the task label that directly affects the input features (RSRPs at primary carrier) and therefore aggregating the dataset from diverse environments helps to learn a better general model as this mapping of UE-to-BS distance and RSRP is not heavily influenced by the environment. For other three tasks, mapping of input features to task labels serve as proxy of environment and local context is hard coded in this relation. Therefore tasks labels are highly influenced by the local context (e.g., layout of buildings within a particular city will affect line of sight and indoor probability as well as secondary carrier strength). Therefore for these three tasks, site-specific model outperforms as compared to aggregating the data samples from diverse cities and base stations for training a general model that hurts the model performance.

\begin{figure}[tb]
	\centering
	\includegraphics[height=0.21\paperheight]{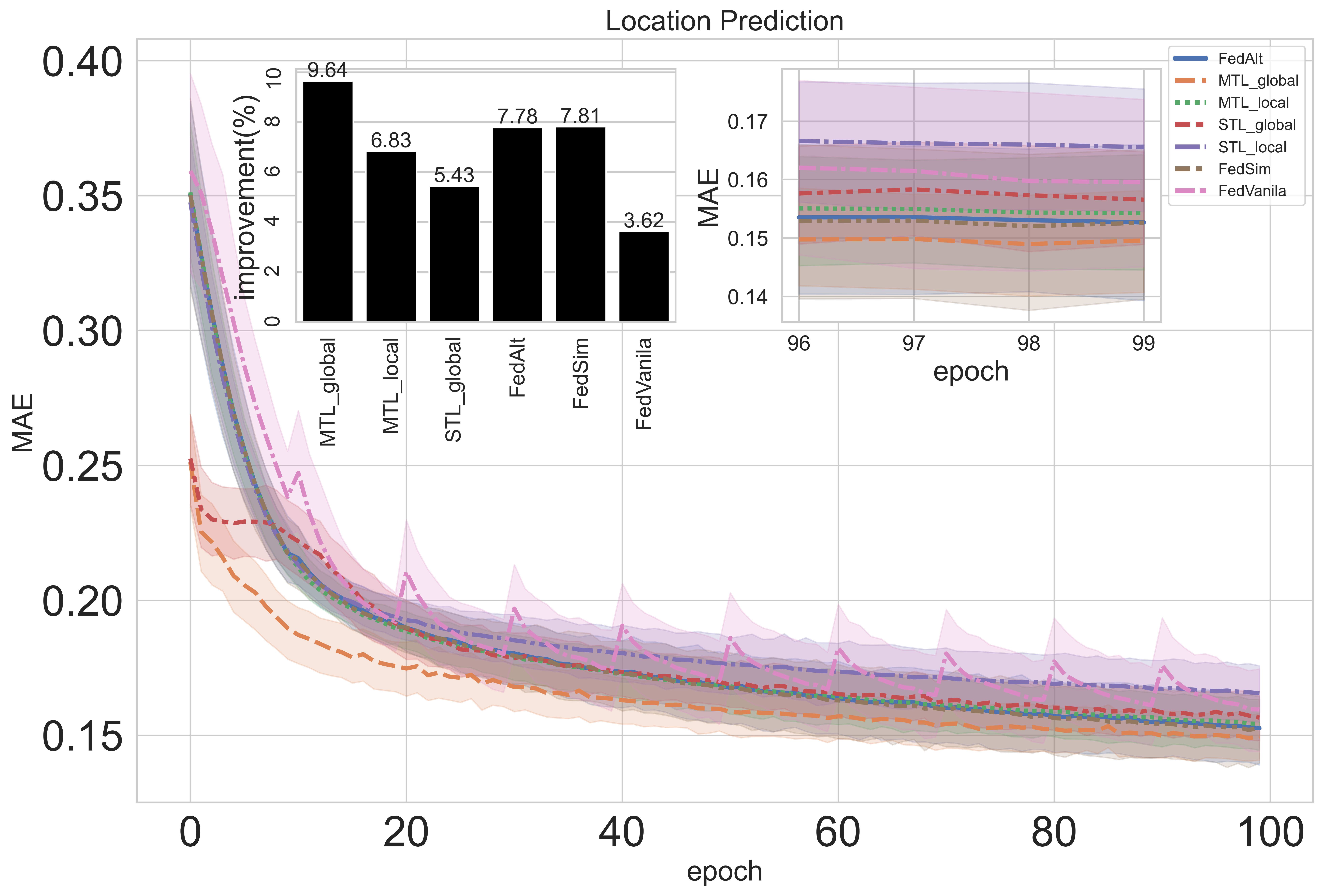}
	\caption{Location Prediction task performance. Improvement over STL\_local = \{MTL\_global: 9.64\%, MTL\_local: 6.83\%, STL\_global: 5.43\%, FedAlt: 7.78\%, FedSim: 7.81\%, FedVanila: 3.62\%\}}
	\vspace{-5 mm}
\end{figure}

\begin{figure}[tb]
	\centering
	\includegraphics[height=0.21\paperheight]{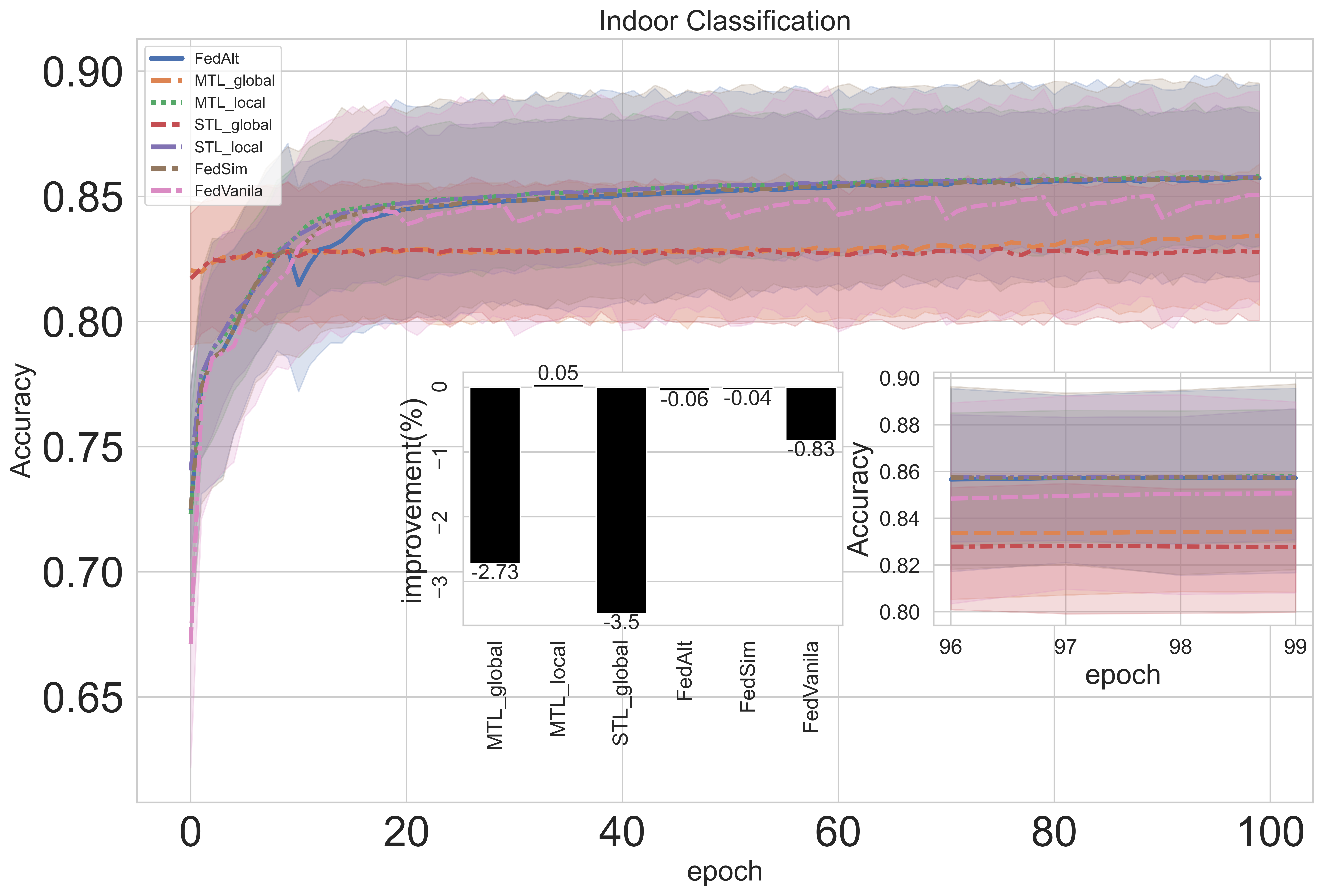}
	\caption{Indoor link classification task performance. Improvement over STL\_local = \{MTL\_global: -2.73\%, MTL\_local: 0.05\%, STL\_global: -3.5\%, FedAlt: -0.06\%, FedSim: -0.04\%, FedVanila: -0.83\%\}}
	\vspace{-5 mm}
\end{figure}

\begin{figure}[tb]
	\centering
	\includegraphics[height=0.21\paperheight]{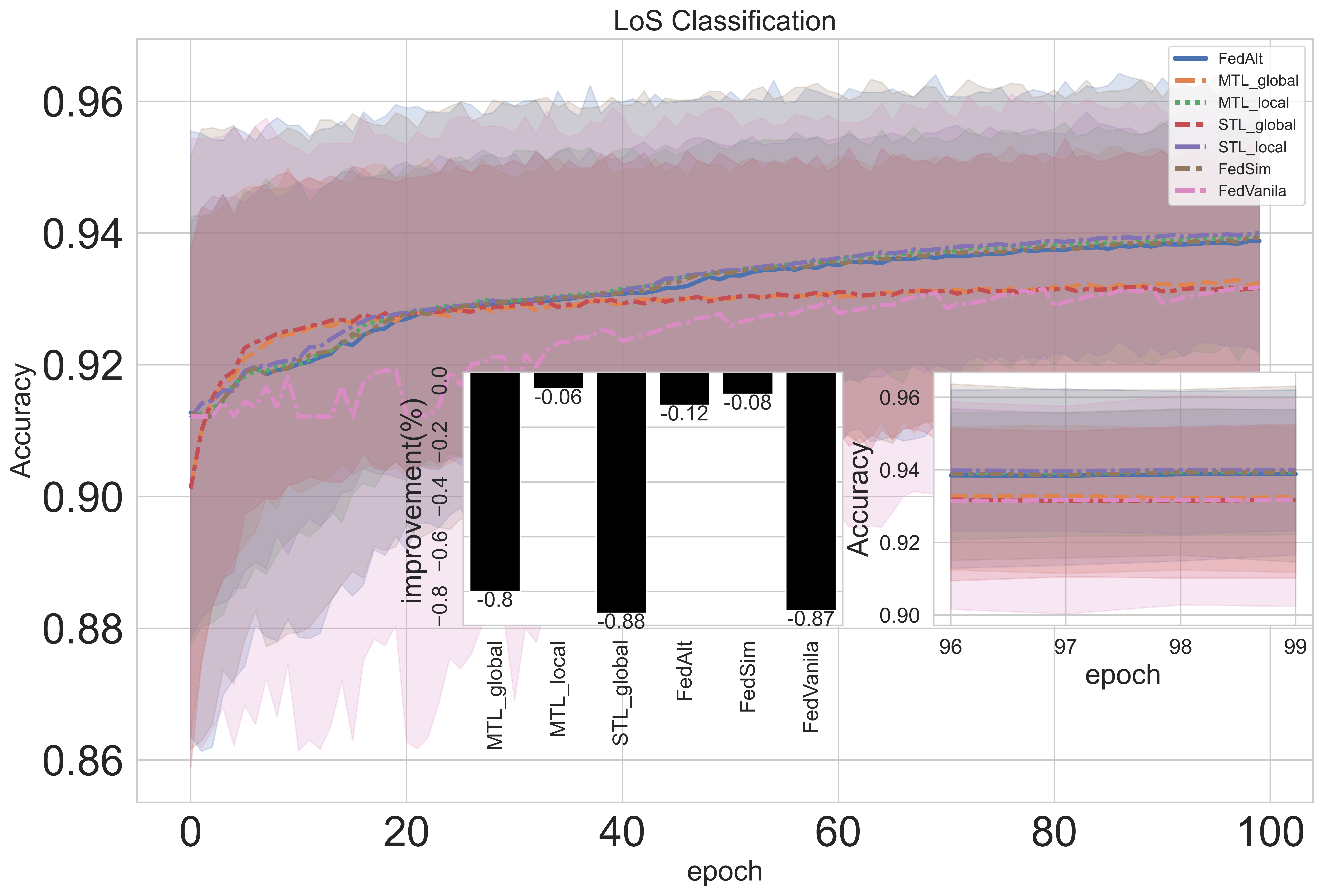}
	\caption{LoS link classification task performance. Improvement over STL\_local = \{MTL\_global: -0.8\%, MTL\_local: -0.06\%, STL\_global: -0.88\%, FedAlt: -0.12\%, FedSim: -0.08\%, FedVanila: -0.87\%\}}
	\vspace{-5 mm}
\end{figure}

Next we experimented with 1\% training dataset size. For space brevity, we have only included visualization for SC task. For secondary carrier prediction (Fig. 9), both globals have quickest convergence by a large margin and outperform all at the end (around 59\% improvement as compared to STL\_local). FedVanila is worst. FedSim is better than FedAlt and FedVanila.  Difference between global and local in STL is almost same as for MTL. All federation based methods fall behind the locals and globals. For location prediction, overall trend was similar with MTL\_global outperforming all (27.3\% improvement as compared to STL\_local) followed very closely by STL\_global and both have very quick convergence. For indoor classification, both globals are better than locals (11.8\% improvement). For LoS link classification, both locals are better than both globals. Both globals and FedVanila perform worse than all while others are on par with each other. Overall with small training dataset availability, globals perform better than locals except in LOS where MTL is on par with STL. For first three tasks, either training a single global model or independent local models is best while for LOS local mode is best. When compared to full dataset size trained model, the performance is worse with models trained on sparse data as can be seen by comparing y-axis of Fig. 5 and Fig. 9.

\begin{figure}[tb]
	\centering
	\includegraphics[height=0.21\paperheight]{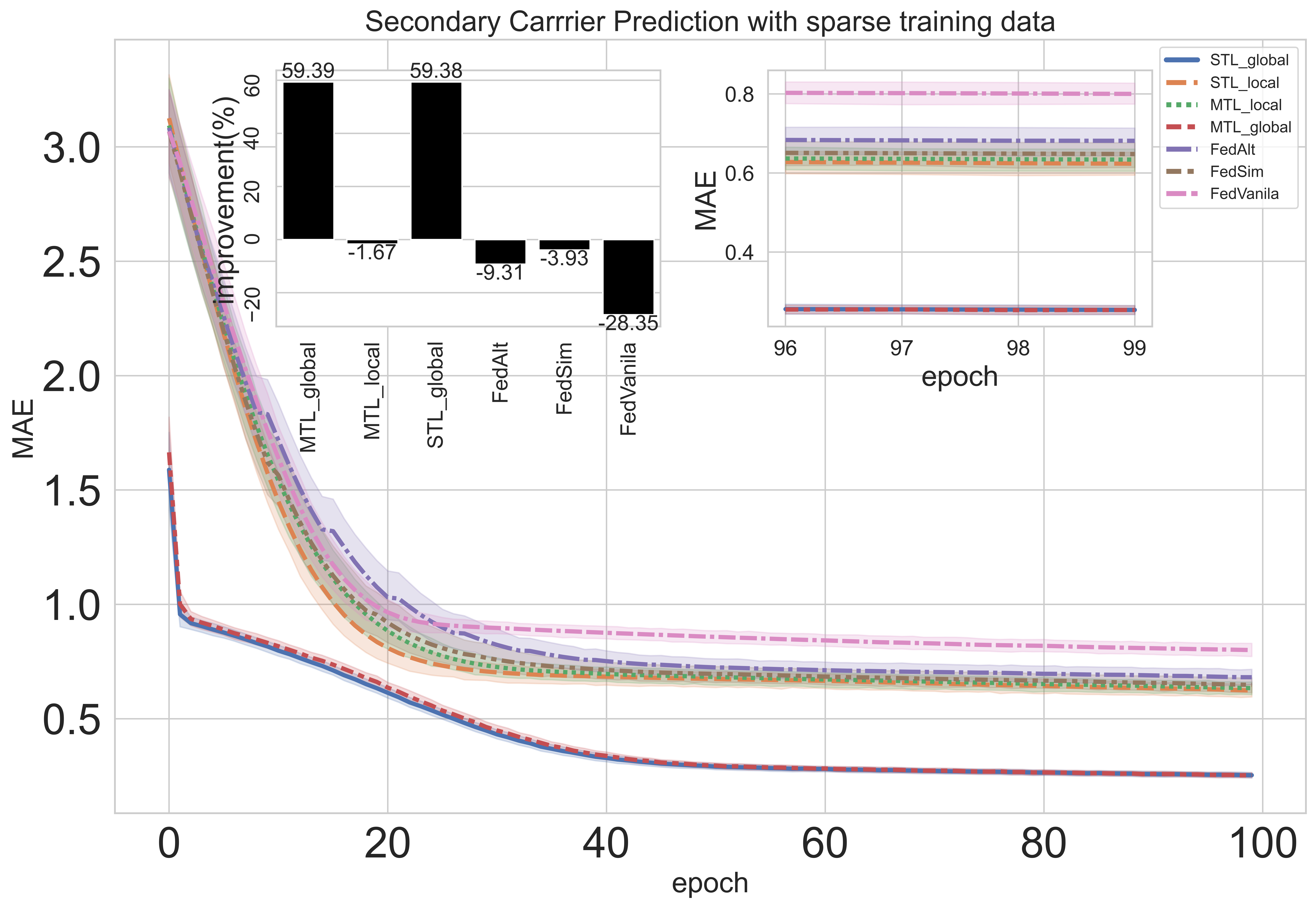}
	\caption{Secondary carrier prediction task performance with sparse data. Improvement over STL\_local = \{MTL\_global: 59.39\%, MTL\_local: -1.67\%, STL\_global: -59.38\%, FedAlt: -9.31\%, FedSim: -3.93\%, FedVanila: -28.35\%\}}
	\vspace{-7 mm}
\end{figure}

Lastly, we analyzed the effect of grouping of tasks and their effect on MTL learning performances using CGC-UW combination and keeping distributed learning fixed to local mode. The convergence were almost on par with each other. Overall,
\begin{enumerate}
	\item 
	 For SC: grouping with PS and LOS outperform all (2.09\% improvement as compared to STL) and second best is with PS (2.06\%). Just having SC as STL is worst.
	 \item
	For PS: grouping with IN outperform all (8.02\% improvement) and second best is with SC (7.38\%). Just PS is worst.	\
	\item
	For IN: grouping with all is best (0.09\% improvement) and second best is grouping with SC, PS and IN (0.08\%). Just IN is worst.
	\item
	For LOS: having it as STL outperformed all and followed very closely by grouping it with SC and LOS. Grouping together with PS and LOS is worst.
\end{enumerate}

In addition to loss/accuracy metric, model sizes and communication costs are also important. MTL using CGC has 62650 trainable parameters while MTL using HPS has much less (16406) trainable parameters. For comparison, STL has 9737, 6657, 5633 and 9737 trainable parameters for the four RAN tasks respectively. Communication cost is highest in global mode then comes FedVanila followed by FedSim and FedAlt and least with local mode.

\section{Conclusions and Future Research Directions}\label{sec:conclusion}

In this work, we have benchmarked the performance of state-of-the-art multi-task learning (MTL) approaches applied to RAN related tasks. As per results, MTL was found to perform either best among all or on par with STL. SC, PS and IN tasks benefited from knowledge sharing (global model or partial federation) while LOS performance degraded. With sparse training data, collaboration through dataset aggregation for a single global model helped significantly but MTL was found to be on par with STL.  Federation with partial parameters is much better than full federation with MTL model. The gain of MTL can be further analyzed by devising a utility value based on model performance and resource utilization costs. For future work, we will investigate scenarios where UEs collect some of the labels of these tasks at the same spatiotemporal point and so labels for all tasks are not available as well as its generalization ability for unseen tasks. We will also investigate distributed algorithms to search for optimal MTL related hyper-parameters of edge nodes (e.g., MTL architecture per node, weighting strategy per node, task grouping per node, number of experts, learning rate etc).

\vspace{-2 mm}

\bibliographystyle{IEEEtran}
\bibliography{mtl}

\end{document}